\documentclass{article}
\usepackage{hyperref, url}
\usepackage{amscd,amsmath,amsthm,amssymb}
\usepackage{enumerate,varioref}
\usepackage{epsfig}
\usepackage{graphicx}
\usepackage{subfig}
\usepackage{mathtools}
\usepackage{tikz}
\usepackage{algorithmicx}
\usepackage{algorithm}
\usepackage{algpseudocode}
\usepackage{xcolor}
\newtheorem{thm}{Theorem}

\theoremstyle{definition}
\newtheorem{defn}[thm]{Definition}
\theoremstyle{remark}

\newtheorem*{rem}{Remark}

\newcommand{\C}{\mathbb{C}}

\title{How to Teach a Quantum Computer a Probability Distribution}
\author{Clark Alexander\\ email: \href{mailto:gcalexander1981@gmail.com}{the author}}

\begin{document}
	
	\maketitle
	
	\begin{abstract}
		Currently there are three major paradigms of quantum computation, the gate model, annealing, and walks on graphs.  The gate model and quantum walks on graphs are universal computation models, while annealing plays within a specific subset of scientific and numerical computations.  Quantum walks on graphs have, however, not received such widespread attention and thus the door is wide open for new applications and algorithms to emerge.  In this paper we explore teaching a coined discrete time quantum walk on a regular graph a probability distribution.  We go through this exercise in two ways.  First we adjust the angles in the maximal torus $\mathbb{T}^d$ where $d$ is the regularity of the graph.  Second, we adjust the parameters of the basis of the Lie algebra $\mathfrak{su}(d)$.  We also discuss some hardware and software concerns as well as immediate applications and the several connections to machine learning.
	\end{abstract}

	\tableofcontents
	
	\section{Introduction}
	In the history of mathematics and computation, the mathematics has often been (far) ahead of the engineering.  In some sense, this is necessary as one won't always know what sort of machine will fit one's needs for specific computations.  Currently (as of 2021) this is certainly the case in quantum computation.  There are, at present, at least three well-known paradigms of computation using quantum mechanical effects.
	\begin{itemize}
		\item The gate model which in some sense attempts to analogize classical computations from a quantum mechanical framework.  This was suggested as early as 1980 \cite{Manin}
		\item Quantum annealing which mimics classical simulated annealing by leveraging quantum tunneling instead of ``temperature" to jump out of local minima.  Classical simulated annealing has been around for quite some time as it is in the suite of Markov chain Monte Carlo methods, while quantum annealing has been around since at least the late 1980s \cite{FACB}.  While quantum annealing is not yet known to be a universal method of computation, it is well suited for optimization problems and problems which rely on random sampling.  Not universal, but still a wide berth of problems and techniques to tackle.  Additionally, since quantum annealing is more specific its mathematics and engineering are further along than the gate model.
		\item Random walks on graphs have found their applications in most fields of modern science and economics.  In the early 2000s, the ``quantization" of random walks lead us to several different paradigms of quantum random walks on graphs, Continuous, Discrete, Coined, and Real \cite{HM, DB, PRR,VBB}.  However, in 2009 Quantum Walks on Graphs were shown to be a universal method of computation \cite{C}.  Despite this, the economic investment in quantum computation by walks on graphs has remained minimal in comparison to the other two paradigms.
	\end{itemize}
	
	In addition to these well-known paradigms, there are quantum inspired algorithms, noisy intermediate scale computers, hybridized algorithms, and heuristic algorithms.  Among the effects quantum computing has on the world in a real sense is that it has sparked a bit of an arms race to see who can produce a faster probabilistic program, and who can leverage quantum effects without needing the sophisticated hardware that is a ``quantum computer."  Because of this, the true natures of the complexity classes \textbf{BQP} and \textbf{BPP} are begin slowly revealed.

	In this work we seek to add a technique into the canon of quantum walks on graphs.  In this work we consider coined discrete time quantum walks on regular graphs.  In this scenario we simultaneously consider two Hilbert spaces, $\mathcal{H}_\mathcal{P}$ (the position space) and $\mathcal{H}_C$ (the coin space).  One can consider a coined walk on a graph which is not regular, but the computation is far less efficient.  In our case the term ``coin" is an analogy to the classical random walk on a line wherein the walker flips a coin and moves accoding to which face of the coin is showing.  In the situation of a regular graph, each vertex has $d$ edges from which to choose, and thus the quantum ``walker" flips a $d$ sided quantum ``coin."  The quantum walker does not move, but instead allows the wave function to propagate across the graph.  The computational efficiency of this approach comes from the fact that we can encode the dynamics of the system in a (complex) matrix of size $Nd \times Nd$.  More important is that this matrix is also unitary.  So our goal the becomes finding a discrete path in $U(Nd)$. 
	
	This work is split in the following way.  In \S1 we discuss some of the technicalities of a discrete time quantum walk on a regular graph as well as some graph properties necessary to ensure our walk will converge to the desired probability distribution.  We also discuss the parameters we can adjust in a DTQW, which are plentiful, and which 
	In \S2 we approach the problem of learning by using the Maximal Torus in $U(d)$ to parameterize our walking space.  We modify the classical algorithm of gradient descent to find our appropriate coins.  It should be noted, that gradient descent is still gradient descent, but we give a very simple implementation by hand which should be understandable to any student who has taken the first semester of calculus and paid attention through the first two weeks of vector calculus.  We approach the problem of trying to fit our target distribution with a single coin, which leaves us with an extremely under determined system and allows us to understand how much we can leverage wave interference to our advantage. In\S3 we use the basis of the Lie algebra $\mathfrak{su}(d)$ which is $d^2-1$ dimensional, thus we essentially square our number of generators.  While for large graphs, and especially ones with good spectral expansion (i.e. the bucky ball, 60 vertices, 3 regular) this is still drastically under determined, we have a better shot and landing on our target distribution.  Consider, for example, a 100 vertex, 10 regular graph.  The quantum walk will have 99 basis elements, for fitting a 100 point vector, so chances are quite good that we can properly fit our distribution. In \S4 and \S5 we give the results and motive future work.  In the end, this particular technique is not extremely useful on its own.  Its utility comes from the fact that we know that there is a plethora of applications for learning a distribution.

	\section{Some Technicalities of Discrete Time Quantum Walks}
	For the sake of consistency, we will use 1-indexing throughout so that our sums appear as $\sum_{i=1}^n$ rather than $\sum_{i=0}^{n-1}$.  The necessary adjustments for 0-indexed languages have been made by the author in the coded implementation of this algorithm and one just needs to pay careful attention throughout to avoid unnecessary headaches with indexing.
	
	A DTQW on a regular graph is an orbit of a unit vector in $\C^{N}$ via a unitary matrix.  The dimension $N$ is the product of the dimension of two smaller Hilbert spaces.  
	\begin{defn}
	Let $G=(V,E)$ be a regular graph with regularity $d$.  Then the two Hilbert spaces one must consider for a DTQW on $G$ are the \emph{position space} $\mathcal{H}_\mathcal{P}$ and the \emph{coin space} $\mathcal{H}_C$.
	\begin{eqnarray}
	\mathcal{H}_\mathcal{P} &=& \text{span} \{ |1 \rangle, \dots ||V|\rangle\}\\
	\nonumber \mathcal{H}_C &=& \text{span} \{|1\rangle,\dots, |d\rangle \}
	\end{eqnarray}
	\end{defn}	
	
	Thus our space $\C^N = \C^{d|V|}$.  We construct a unitary 
	\[
	\hat{U} : \mathcal{H}_C \otimes \mathcal{H}_\mathcal{P} \rightarrow \mathcal{H}_C \otimes \mathcal{H}_\mathcal{P}
	\]
	as the product of two other unitaries $\hat{S}$ and $\hat{C}$.
	
	\begin{defn}
		The two unitary operator $\hat{S}$ and $\hat{C}$ are defined as follows.  For the ``coin operator" $\hat{C}$ we first pick a unitary matrix $C_0$ from $U(d)$ which could be at random or through a parametrization
		and we define
		\begin{equation}
		\hat{C} = C_0 \otimes I_{|V|}
		\end{equation}
		The operator $\hat{S}$ called the ``shifting operator," is built by analogy to a classical random walk.  In a classical random walk one flips a coin, and then moves accordingly.  In the quantum walk, one projects onto a ``con flip" and moves according the the graph structure.
		\begin{equation}
		\hat{S} = \sum_{j=1}^d \sum_{v\in V} |c_j\rangle \langle c_j| \otimes |w\rangle \langle v|
		\end{equation}
		Where $w$ is the $j^{th}$ adjacency to $v$.
	\end{defn}
	\begin{rem}
		One must be careful to respect the graph structure so that $\hat{S}$ remains unitary.  This is done by reading the adjacencies from a consistently labeled rotation map cf\cite{A}
	\end{rem}

	Our discrete time quantum walk can now succinctly written as
	\[
	|\psi_t\rangle = \hat{U}^t |\psi_0\rangle 
	\]
	
	The to compute the probability that our quantum walker will collapse into a classical walker at vertex $i$ we have
	\begin{equation}
	Prob(\psi_t \text{ is at vertex } i) = \sum_{j=1}^d |\langle c_j|\otimes \langle i|\psi)t\rangle|^2
	\end{equation}
	
	Now if one is observing closely, there are a lot of parameters in play here.  Let's list some of them
	\begin{itemize}
		\item In $\hat{C}$ we have a choice of any unitary matrix.  That gives us $d^2-1$ dimensions of freedom
		\item In $\hat{S}$ we have a  choice of a consistent edge labeling of our graph.  These are not unique, in fact for graphs with a high degree of regularity the number of consistent edge labeling grows algebraically (tight bounds on this growth are not currently known to the author)
		\item We have a choice of starting position.  Generally this is taken as $|1\rangle_{\mathcal{P}}$. This is not necessary, the starting position can be any unit length vector in $\C^{|V|}$.
		\item We have a choice of a starting coin.  This is generally taken as the \emph{Fourier coin}
		\begin{equation}
		|c\rangle = \sum_j e^{2\pi i(j-1)/d}|j\rangle_C
		\end{equation}
		Again, this is not required, it is just a convention, giving us another $d-1$ dimensions of freedom.
		\item Finally, we have the evolution time of $\hat{U}$.  This can be any integer, positive or negative.
	\end{itemize}

    With this many degrees of freedom, it's almost a wonder the DTQW have any utility.  But this is where the difficult mathematics comes in.  For our purposes we will set the number of steps in our evolution ahead of time.  This will be set somewhere near $2*$ diameter$(G)$.  We will also keep the conventions of starting at the first vertex in the graph with a Fourier coin.  While the shifting operator does have some degree of freedom, we will choose a rather simple edge labeling.  In fact, we will choose the first consistent edge labeling we can solve.  As long as we continue with this edge labeling, the operators will remain unitary.  This leaves us only with coin.  Therefore for this work we will stick to ``training" the coin to produce the quantum walk we want.

    One other thing of which we must be aware is that in the situation of a bipartite graph, one cannot guarantee certain distributions.  Therefore we have two potential fixes for this.  The first is to make sure our graph has an odd length cycle.  If this is not available, then we must start with our quantum walker in a superposition of each part of the bipartite graph.  Consider the square ($C_4$) for example.  If we start at $|1\rangle$ and evolve for an odd number of steps, we will not land on $|1\rangle$.  Thus we may start with the distribution $(|1\rangle + |2\rangle)/\sqrt{2}$ or with the uniform distribution.

	\section{Method One: Learning the Angles in a Torus}
	\subsection{A Modified Gradient Descent}
	For the coin operator in DTQW we need to choose some parametrization.  Since the coin operator is of the form $C\otimes I_{|V|}$ we only need to choose a parametrization of the space $U(d)$.  This parametrization need not be topologically dense, but it must simply allow us to move around ``enough" in the space. For the case of a single coin, we have chosen to parametrize the maximal torus of $U(d)$.  This alone does not, however, make a good quantum walking matrix as it is a diagonal matrix and will simply produce a phase shift at each vertex without affecting the probabilities.  On the other hand, the Fourier matrix tends to disperse waves quite well, but is itself only an order 4 operator ($\mathcal{F}^4 = I$) thus we have chosen the conjugate an element of the maximal torus by a Fourier Matrix
	\begin{equation}
	C = \mathcal{F}^* \begin{bmatrix}
	e^{i\theta_1} & & &\\
	& e^{i\theta_2} & &\\
	& & \ddots & \\
	& & & e^{i\theta_d}
	\end{bmatrix} \mathcal{F}
	\end{equation}
	
	This leaves us with a very tiny amount of parameters to optimize. We shall consider our parameter set as a vector $\vec{\theta}$ and we define the gradient descent algorithm as normal
	\begin{equation}
	\vec{\theta}_{n+1} = \vec{\theta}_n - \gamma \nabla_{\theta} E(\vec{\theta}_n)
	\end{equation} 
	
	where $\gamma$ is the learning rate and $E$ is some error dependent on $\vec{\theta}$.
	
	None of this is terribly surprising.  Where things get difficult is in actually computing the gradient.  Since we are trying to learn a probability distribution $\vec{\pi}$ we define the error as
	\begin{equation}
	E(\vec{\theta}) = \left(\sum (|\langle j | \hat{U}^t |\psi_0\rangle|^2 - \pi_j)^2\right)^{1/2}
	\end{equation}

	Again nothing terribly surprising.  The difficulty is in trying to take the derivative of this expression with respect to $\theta_j$
	
	\begin{equation}
	\frac{\partial E}{\partial \theta_j} = \text{ a big mess }
	\end{equation}
	
	This computation is feasible by hand (as the author has performed it several times), but computationally it's not worth the effort for the mathematical rigor and precision.  One must remember that we are trying to approximate the angles with floats to within some tolerance.  Computers cannot handle mathematical exactness in memory, however, computers can approximate extremely fast, and so we have chosen to take advantage of speed in both simplicitiy of code and ease of reading. We set some tolerance $\varepsilon = 0.01$ and we compute
	\begin{equation}
	\frac{\partial E}{\partial \theta_j} \approx \frac{E(\vec{\theta} + \varepsilon \hat{e}_j) - E(\vec{\theta} - \varepsilon \hat{e}_j)}{ 2\varepsilon}
	\end{equation}
	
	It is, however, an approximation that we seek, and this gradient can be computed in a few lines of code rather than the several lines it takes to write out the true partial derivatives.  
	\begin{rem}
		One of the several difficulties in computing the true partial derivatives is that when considering a matrix 
		\[
		\hat{U}^t = \hat{S}\hat{C}\cdots \hat{S}\hat{C}
		\]
		The standard Leibniz Rule from Calculus I fails since $[\hat{S},\hat{C}] \ne 0$.  There are higher order noncommutative effects that must be handled carefully.  If we choose evenly a moderately small $t$ (20 for example) then the formaula for the true partial derivative grows in length to several pages handwritten and grows to unreadable in computer code.
	\end{rem}
	
	\subsection{A Single Coin}
	In the method of using a single coin in the maximal torus, we have restricted ourselves to exactly $d+1$ parameters.  The $d$ angles along the torus, plus the number of steps we wish to take.  Classical feed forward neural networks have far more parameters than the desired probability distribution.  So our goal is to control the waves emanating from the first vertex in such a way that they constructively and destructively interfere with each other to land on precisely the probability distribution we wish.  It is in this sense that we are leveraging wave interference to take over some of the parameter tuning.
	
	In \S5 we will look at some results from using a single coin and how preselecting the number of steps in the walk affects the training.  Suffice it to say, using a small number of steps is not sufficient to properly train a DTQW to learn an arbitrary distribution.  One instead needs to solve the equation of picking angles so that $\hat{U}$ has a topologically dense orbit in $\mathbb{S}^{2d|V|-1}$ and then solving the number of steps to get withing a desired tolerance when projecting down to $\mathbb{S}^{2|V|-1}$.

	\subsection{A Single Coin per Step}
	
	The approach that tends to work empirically is allowing oneself to pick a different coin at every step.  Intuitively this makes much more sense.  One moves a little closer to the desired distribution and then needs to change direction slightly so that in the end one form a discrete path in $\mathbb{S}^{2d|V|-1}$
	
	We will see again in \S5 how this approach differs from that of allowing only a single coin.  The technical difficult here is training so many more parameters.  It mertis mention that for graphs with good spectral expansion properties, $d$ is relatively small versus $|V|$ and thus using several steps each with $d$ parameters still goes far beneath a general set up in a feed forward neural network.
	
	Depending on one's computational power, choosing a higher number of steps is preferable as in some experiments the author has seen the last coin or two coins have parameters of 0.  This means, the walk can converge in fewer steps, and the coin matrix, given the current parametrization is simply $I_d\otimes I_{|V|}$.  
	
	\section{Method Two: Learning the Parameters in a Lie Algebra}
	\subsection{A Single Set of Parameters}
	
	As we mentioned in the first method, we have a severely under determined system, but we used only the Maximal Torus which is dimension $d$.  In the second method we parameterize the whole of $\mathfrak{su}(d)$ by building a version of the Gell-Mann matrices.  While properly the Gell-Mann matrices are the basis of $\mathfrak{su}(3)$ one can extend the basis to include $d$ dimensions.  The basis of $\mathfrak{su}(d)$ is $d^2-1$ dimensions and so we have a significant advantage in using a single coin.  In the computions the author did not use the exact Gell-Mann matrices, but a similar set which is easy to build.  Let's take a small detour to give the construction.
	
	\subsubsection[basis matrices]{The Basis Matrices of $\mathfrak{su}(d)$}
	We know that the $\mathfrak{su}(d)$ consists of skew-hermitian matrices.  Defining the matrices 
	\[E_{ij} = \hat{e}_i\hat{e}_j^t\]
	That is, a matrix of zeros with a 1 in row $i$ column $j$.  Then for $i < j$ we have skew-hermitian matrices
	\[
	E_{ij} + E_{ji} \text{ and } -iE_{ij} + iE_{ji}
	\]
	
	This gives us $\binom{d}{2} + \binom{d}{2}$ matrices.  We also have the diagonal matrices $E_{ii} - E_{i+1,i+1}$ where $i$ runs from 1 to $d-1$.  This gives us $2\binom{d}{2} + d-1 = d^2-1$ matrices.  It should be noted that these matrices do not satisfy the Lie algebra structure constants in any canonical way, but they do form a basis.  They are also easier to compute and program.  Labeling these matrices as $\lambda_1, \dots, \lambda_n$  we see that our coin can be written
	\begin{equation}
	C = \exp^{-i\vec{\alpha}\cdot \vec{\lambda}}
	\end{equation}
	
	Thus we seek to learn the $d^2-1$ parameters $\vec{\alpha}$.  This general construction in $U(d)$ allows us to get at some of the results of \cite{VBB} without needing to entangle coins in lower dimensional spaces.  When we consider zig-zag products or even Cartesian Products we automatically increase the dimensionality of our coin Hilbert space.

	As one can imagine this generally gives us more success that training a single coin on the Maximal Torus.  
	
	\subsection{A Single Set of Parameters per Step}
	
    The difference in the method from \S4.1 to \S4.2 is akin to the differences in \S3.2 and \S3.3, however, given the high number of parameters to tune, this method becomes computationally expensive after even a small graph.  For this reason, we show examples only from \S3.3 and omit training a general walk of $n$ steps required $n(d^2-1)$ parameters to train at every iteration.    
	
	\section{Results and Examples}
	\subsection{Petersen Graph; a Single Coin}
	The first graph we will explore is the Petersen graph.  In this case we have a 10-vertex, 3 regular graph, which means our quantum walking matrix is $30\times 30$.  We were unable to fit a randomly drawn distribution with a single coin, nor were we able to with a single set of parameters.  Just to get a sense of this, consider the following graph:
	\[
	\includegraphics{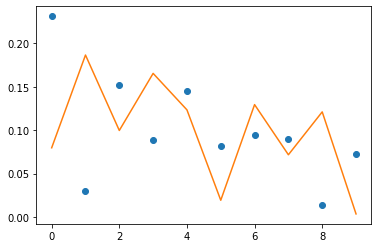}
	\]
	This is a graph of the best fit we could get on the Petersen graph using a single coin and 20 steps.  Notice that we have certainly done better than having all the mass accumulated at vertex 1, but still not really teaching the graph the correct distribution.  We can see, however, that it has taken the parameter of 20 steps seriously.  Consider the following graph of errors per step.
	
	\[
	\includegraphics{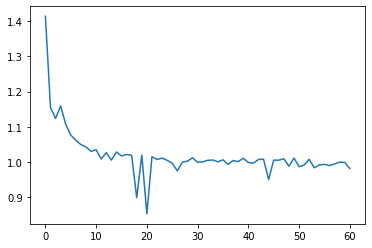}
	\]
	
	We see a clear downward spike at exactly 20.  This means the graph was training to learn how to produce the desired probability using exactly 20 steps.  Part of how we should like to approach quantum image processing is moving one step further or one step less using the ``correct" coinage.
	
	\subsection{Petersen; One Coin per Step}
	Now let's consider using a single coin per step.  In each case we used 8 steps with 3 parameters per step.  These were fit using a completely randomized initial set of coins, and run for 300 iterations or gradient descent.
	
	The first graph is a randomly drawn distribution.  The orange line is the probability distribution, and the blue dots are the learned distribution.
	\[
	\includegraphics{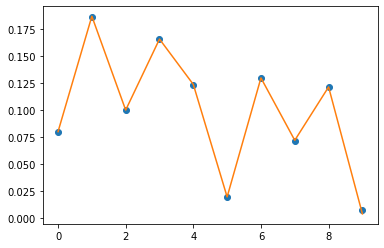}
	\]

	The second graph is the most difficult.  We tried to get a quantum walk to converge to a single point. The difficulty is in getting all the wave function to destructively interfere everywhere except a single vertex, but constructively interfere at exactly one vertex.  After 300 iterations we arrived at the following.
	\[
	\includegraphics{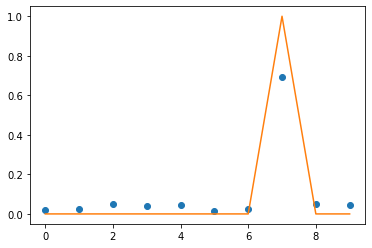}
	\]

	The third distribution we have chosen to train is the Boltzmann distribution.  This was the best fit of the three on the Petersen graph.
	
	\[
	\includegraphics{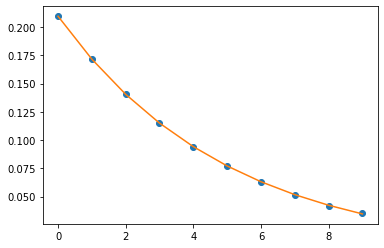}
	\]
	
	\subsection{Fullerene $C_{60}$; One Coin per Step}
	
	Now let's give ourselves a big challenge.  We're going to try to have the fullerene graph $C_{60}$ learn a Gaussian distribution.  To make things even more complicated, let's set a few parameters.  The diameter of $C_{60}$is 9.  So we can't simply use 9 steps.  That's asking too much.  We'll use 13 steps.  There is very little room for error in 13 steps on $C_{60}$.  Additionally, the vertex labeling given in a somewhat random fashion rather than labeling vertices in a Hamiltonian path.  We use the 5-fold symmetric two dimensional projection and begin in the center and label vertices in a counterclockwise direction moving outward.  The Gaussian distribution is centered between vertices 29 and 30, in fact we're using
	\[
	\pi = A\exp(-(x-29.5)^2/(180)) 
	\]
	where $A$ is a normalizing constant.  That is, we're using $\sigma^2 = 3|V|/2$.  This gives a nice symmetric look in the graph and doesn't concentrate the probabilities too much in the center (making the distribution harder to learn).  So without further ado, let's take a look at some partial results.
	We start at vertex 1 with all the probability giving us this graph
	\[
	\includegraphics{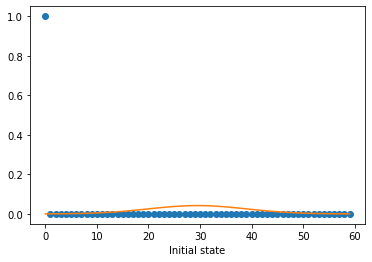}
	\]
	We can check in on the progress at several iterations.
	\[
	\includegraphics{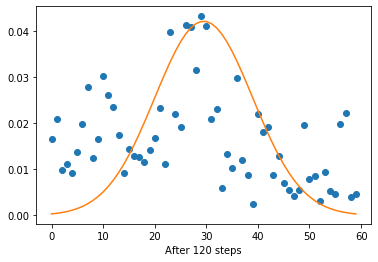}
	\]
	\[
	\includegraphics{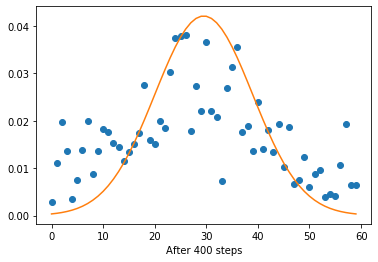}
	\]
	
	The challenge we've given ourselves is immense.  This graph suffers from vanishing gradients quickly.  So after about 1000 iterations, we substituted a mildly stochastic gradient descent.
	
	\[
	\includegraphics{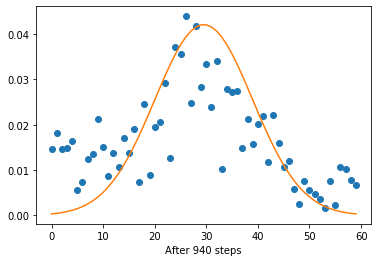}
	\]
	\[
	\includegraphics{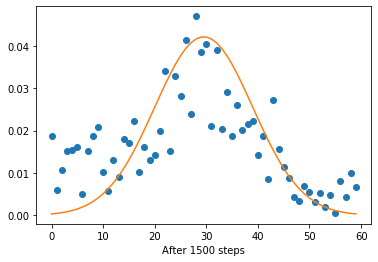}
	\]
	\[
	\includegraphics{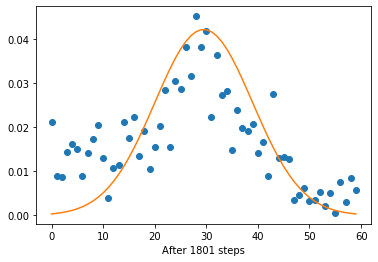}
	\]
	
	We now see the probabilities starting to settle onto the distribution.  Given the rate at which errors decrease, we expect this to settle in around 6000 or so iterations.
	
	
	\[
	\includegraphics{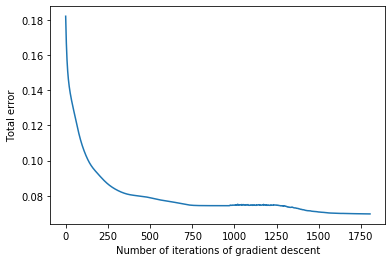}
	\]
	
	We can see the small bump in errors starting around 1000 iterations.  This was a modified stochastic gradient descent.
	
	\subsection{A Picture of Oak Ridge}
	Let's look at one more fun aspect of quantum walking.  Here's a picture of Oak Ridge, TN (not the lab, just some road in autumn)
	\[
	\includegraphics[width = 8cm]{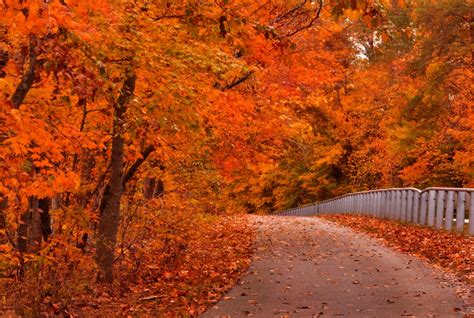}
	\]
	
	Now consider the following related image
	\[
	\includegraphics[width = 8cm]{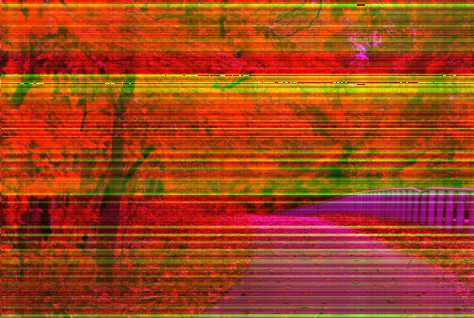}
	\]
	
	This may be the first image ever generated by quantum image processing via discrete time quantum walks on graphs.
	
	The process to create this second image is fairly simple.  
	\begin{enumerate}
		\item Using an image library, separate the channels into red, green, blue.  
		\item For this image we used only the green channel, and separated it into columns.  This image had channels of size $318\times 474$.  That's too large to run an entire quantum walk on a laptop, so we ran this column by column.
		\item For each column reduce the numerical input to a probability mass function $C \rightarrow C/\sum C$
		\item Initialize a discrete time quantum walk on the cyclic graph $C_{318}$, with initial position vector as the probability of the column
		\item Build a walking matrix of two steps with small, but random coin angles.
		\item Measure the probability of the 2 step quantum walk, then rescale.
		\item Replace the original column with the rounded, scaled quantum walk probabilities. 
	\end{enumerate}

	Looking closely at the second image, the wave patterns appear vertically.  This reveals the quantum mechanical nature of the walk, rather than choosing randomly or simply shifting or running a few steps in a Markov chain.
	
	When the hardware is scaled up, the author would like to train a quantum walk on a large probability mass function and run a quantum walk with initial state as the image itself, and advanced it forward.

	\section{Applications}
	
	This work is at the intersection of several exciting topics as of early 2021.  Namely, quantum computing and machine learning.  These, however, are only the broadest brushes with which we can paint this method.  Learning a probability distribution, however, is the goal of several famous Monte Carlo methods, for example, Metropolis and Metropolis-Hastings, and Metropolis Coupled Markov Chain Monte Carlo (MCMCMC)\cite{mc3}, Metropolis-adjusted Langevin Algorithm \cite{RT}, etc.  Recently there has been some movement on learning distributions for time series forecasting\cite{BBO}, hurricane forecasting \cite{WT}, image processing, music generation, and natural language processing. \cite{WDL}  
	
	In the actual implementation of this work, the goal is to teach a quantum walk how to control the angles of its interfering waves.  While this is not directly control theory, it is tangentially related.
	An additional area of application which is receiving some attention in the quantum world is ``post-quantum" cryptography.  In \cite{EHMLP} the authors develop cryptography schemes on super singular isogeny graphs.  These are regular graphs whose vertices represent super-singular elliptic curves over finite fields.  The edges represent morphisms between them.  This gives a family of graphs with $q$ vertices and $q(q-1)/4$ edges for $q\equiv1 \mod 4$.  The graph walking methods introduced here are able to handle many of these graphs on digital computers.  One should note, however, that as of 2021 there is no known quantum algorithm which can break this cryptography scheme.  Perhaps quantum walks on super-singular isogeny graphs could be of some use.  
		
	\section{Open Questions and Future Work}
	
	As this work is relatively early in the intersection of DTQW and machine learning there is a bevy of open questions around it.  Among the most obvious to the author are, how can this be made more efficient.  One practicality in this work is that quantum walks on graphs can be honestly computed  on digital computers if one allows for tiny rounding errors.  However, while attempting to compute a distribution fitting scheme on the cross product of two graphs (a 14-vertex 3-regular with a 7-vertex 4-regular graph) the relatively small size of 98 vertices on a 7-regular graph took an inordinate amount of time.  In using a single coin per step in the maximal torus scheme, only 8 coins causes  one to train 56 parameters per iteration and use a product of fairly sparse matrices of size $686\times 686$.  This is memory intensive and very computationally expensive.  It is not fully clear the role that good spectral expander play in the DTQW universe is yet, however, having a low degree of regularity and small diameter reduces the number of parameters one needs to train, but increases the sparsity of the matrices involved.  Take the buckyball for example.  It's 60-vertex 3-regular structure means that the adjacency matrix has density 180/3600 = 0.05.  Exactly 95\% of the matrix is empty.  It's diameter is 9.  So we need at least 9 coins to even traverse the entire graph.  
	
	A second open question that has come up several times in the course of this research is the existence and implementation of a polynomial time constructive algorithm for consistent roation maps for regular graphs.  \cite{A} shows the necessity of having such maps for ``efficient" construction of DTQW.
	
	A third question we'd like to explore is the roll of directed regular graphs, and their slight generalization of weighted graphs,for example \cite{wong}.  Weighted graphs have found incredibly many uses in machine learning an optimizations.  With the aid of \cite{S} one expects quantum speed ups of Markov Chain algorithms, but the question becomes what is the necessary hardware to implement an algorithm with any utility outside of pure academic interest.  In particular, the utility of a consistent rotation map on regular graphs reduces the computational space of DTQW from $|V|^2 \times |V|^2$ to $d|V|\times d|V|$ which reduces computational cost by density$^{-2}$.  How far can we take this.
	
	A fourth question we are considering and working in actively, is the successful implementation of real quantum walks on regular graphs.  This will improve the size of problems that we can tackle since some modern numerical packages (in particular numpy) has a terrible difficulty with complex numbers.  By reducing our coin space from $U(d)$ to $O(d)$ or even $SO(d)$ we cut our dimensions in half.  That is $mathbb{S}^{2d|V|-1} \rightarrow \mathbb{S}^{d|V|}$.
	
	Another area the author is actively exploring is hybridizing feed forward neural networks by quantum hidden layers.  The quantum hidden layers are multiple random quantum walkers on a graph in between classical Dense layers in a feed forward network.  
	
	A final question and area of future work is the rigorous calculation of material properties using DTQW or CTQW on molecular graphs.

	\subsection{Proposed Additional Methods for Finding Angles}
	
	In order to increase the efficiency of solving the coins for a quantum walk we propose using several additional methods.  The first method we propose is for the maximal torus angle finding.  Instead of using a diagonal matrix in form of 
	\[
	\begin{bmatrix}
	e^{i\theta_1} & & &\\
	& e^{i\theta_2} & &\\
	& & \ddots & \\
	& & & e^{i\theta_d}
	\end{bmatrix}
	\]
	
	which is $2\pi$ periodic to a matrix with unit periodicities 
	
	\[
	\begin{bmatrix}
	e^{2\pi i\theta_1} & & &\\
	& e^{2\pi i\theta_2} & &\\
	& & \ddots & \\
	& & & e^{2\pi i\theta_d}
	\end{bmatrix}
	\]
	
	Then we can use either a classic Monte Carlo random sampling, or in lower dimensions, the quasi-Monte Carlo method of sampling by low discrepancy sequences.  In the case of cubic graphs, we have bounds for optimal low discrepancy sequences in $[0,1]^3$. \cite{WLDS}
	
	We need not use low discrepancy sequences to find the angles exactly, but there is essentially no computation aside from the guess and check method.  The lowest error derived by a low discrepancy sequence can be the initial angles in our coin, thus significantly reducing the number of iterations required in gradient descent to land on our desired probability distribution.
	
	A second proposed method is to use an evolutionary algorithm, such as a genetic algorithm, simulated annealing, or other MCMC algorithm modified for optimization.  From the memory perspective we will almost always prefer simulated annealing to a genetic algorithm, but for ease of implementation a genetic algorithm will win.  
	
	The basic idea of implementation is to choose our angles at random, measure the error and then guess and measure again.  Whichever is better, we keep going in its direction.  We allow ourselves to step backward from time to time, but that is determined by a parameter (often called temperature).  To change our guess, we draw from a proposed probability distribution (often the normal or uniform distributions, although we could use a DTQW here to draw random samples).  The efficacy of this method comes when we draw small samples and add them to the current samples.  That is
	\[
	f(\vec{x} + \vec{\varepsilon}) \approx f(\vec{x}) + \nabla_x f(\vec{x})\cdot \vec{\varepsilon}
	\]
	which is, of course, a simulated gradient calculation.  
	\section*{Appendix: Where to Find the Code}
	The author will periodically release modules of this code to GitHub, however, a better bet is to send a direct communication to the author and a small zip can be sent via email.  Languages are primarily Python and Julia, although Octave can still be found in some places where numerical tests were run.

\end{document}